# Asymptotic expansion based equation of state for hard-disk fluids offering accurate virial coefficients

Jianxiang Tian [1, 2, 4], Yuanxing Gui [2], A. Mulero [3]

[1]Shandong Provincial Key Laboratory of Laser Polarization and Information Technology

Department of Physics, Qufu Normal University, Qufu, 273165, P. R. China

[2]Department of Physics, Dalian University of Technology, Dalian, 116024, P. R. China

[3]Department of Applied Physics, University of Extremadura, Badajoz 06072, Spain

[4]Corresponding author, E-mail address: jxtian@dlut.edu.cn

## Abstract:

Despite the fact that more that more than 30 analytical expressions for the equation of state of hard-disk fluids have been proposed in the literature, none of them is capable of reproducing the currently accepted numeric or estimated values for the first eighteen virial coefficients.

Using the asymptotic expansion method, extended to the first ten virial coefficients for hard-disk fluids, fifty-seven new expressions for the equation of state have been studied. Of these, a new equation of state is selected which reproduces accurately all the first eighteen virial coefficients. Comparisons for the compressibility factor with computer simulations show that this new equation is as accurate as other similar expressions with the same number of parameters. Finally, the location of the poles of the 57 new equations shows that there are some particular configurations which could give both the accurate virial coefficients and the correct closest packing fraction in the future when higher virial coefficients than the tenth are numerically calculated.

## 1. Introduction

As is well known, the hard-disk (HD) fluid is defined by an interaction potential that considers only the repulsive forces among two-dimensional molecules.[1] The simplicity of this model permits it to be used as reference in the description of some phenomena in surface science such as the prediction of monolayer adsorption isotherms.[2-4]

Unfortunately, there is no exact theoretical solution for the equation of state (EOS) of this two-dimensional system.[1] As a consequence, a great variety of analytical expressions for the HD EOS have been proposed over the course of many years, a situation which continues today as from time to time new equations of state are proposed for this fluid.[1, 5-15]

Recently, an extensive review including more than thirty analytical expressions for the HD EOS has been published,[13] and the current state-of-the-art has been summarized in Ref. 14. In particular, it was shown that the most accurate HD EOSs with respect to reproducing computer simulation data for the compressibility factor, $Z$, are those proposed by Kolafa and Rottner,[10] which contain 12 or 13 parameters obtained from both the first five virial coefficients[11] and a fit to their own computer simulation results.

With respect to EOSs reproducing the exact or numeric values of the first ten virial coefficients within their respective uncertainties, five expressions are available: the two Padé equations proposed by Clisby and McCoy (CM),[11] constructed using those first ten values, and the three Kolafa and Rottner (KR)[10] EOSs. Nevertheless, the CM expressions do not give accurate values for $Z$ at very high densities, and the KR expressions do not closely approximate the numeric CM values for the virial coefficients. For this reason, Mulero et al.[14] proposed the so-called SMC1 EOS which contains 13 parameters obtained by using both the values of the first ten virial coefficients from Ref. 11 and a fit to the computer simulation results of Ref. 10. The SMC1 EOS reproduces the values of the aforementioned coefficients even better than other similar approximants, and gives excellent accuracy in the reproduction of computer simulation values of $Z$ over the whole density range, although not as accurate as the KR EOSs. The prediction of virial coefficients higher than the tenth was not considered there.

More recently, Santos and López de Haro[15] have proposed a new kind of analytical EOSs for hard disks, spheres, and hyper-spheres. These are called branch-point approximants (BPA), and contain seven parameters that are obtained only from the first seven virial coefficients values, i.e., without a fit to computer simulation data for the compressibility factor. Due to the simplicity of the analytical expression, the values this new EOS gives for the higher virial coefficients are only approximate.

Taking all these previous results into account, and with Clisby and McCoy [11] having made estimates of the values of the 11th to the 18th virial coefficients for HDs, it was interesting to ask whether an analytical EOS constructed using only the values for the first ten virial coefficients, and different from the CM EOSs, could reproduce those estimates.

As we have recently shown,[16] the asymptotic expansion method (AEM) proposed by Khanpour and Parsafar [12] is well-suited to constructing an accurate EOS using only numeric virial coefficient values, *i.e.*, without using a fit to computer simulation data for the compressibility factor. We must note also that we extended the AEM in order to consider both positive and negative integer exponents in the expansion terms.[16]

The present work studies the performance and accuracy of the same method used for hard spheres in Ref. 16 but now in the two-dimensional case. Three expressions out of 57 studied were first selected, and their accuracy tested by comparing the virial coefficients obtained with the currently accepted numeric values from Ref. 11, and the compressibility factor values with computer simulation data.[10] Comparison was also made with some other recent analytical HD EOSs.

The paper is organized as follows: in Sec. 2, we review the current state-of-the-art about the knowledge of virial coefficients and new equations of state for hard-disk fluids; in Sec. 3 the extended asymptotic expansion method is used to construct the new equations of state; in Sec. 4, we present and discuss our results; and Sec. 5 gives the concluding remarks and prospects for future developments.

## 2. Virial Coefficients and Equations of State

Virial coefficients can be considered to be the cornerstones of the theory of fluids.[17] They are the coefficients in the density expansion of the EOS expressed via the compressibility factor, $Z$, as follows:

$$Z = \frac{P}{\rho k_B T} = 1 + \sum_{i=2}^{\infty} B_i y^{i-1} \tag{1}$$

where $P$ is the pressure, $\rho$ the reduced surface density, $T$ the temperature, $k_B$ Boltzmann's constant, and the packing fraction, $y$, is defined in HD fluids as $y = \pi\rho/4$.

The virial coefficients $B_i$ are defined by exact formulas in terms of integrals whose integrands are products of Mayer functions. For hard disks, those integrals are numbers (they do not depend on temperature), but unfortunately only the first four can be calculated analytically [17]:

$$B_2 = 2; \quad B_3/B_2^2 = 4/3 - \sqrt{3}/\pi = 0.782004...;$$

$$B_4/B_2^3 = 2 - 4.5\sqrt{3}/\pi + 10/\pi^2 = 0.5322318... \tag{2}$$

and then can be considered as exact values.

Values for the virial coefficients from the fifth to the tenth have been obtained by numeric integration by Clisby and McCoy,[11] who also made estimates of the eleventh to the eighteenth ones. These estimated values are given in the first column of Table 1, where the uncertainty is in the last decimal digit.

Although the convergence of the virial series, Eq. (1), is not guaranteed for all state points, the fact is that the more virial coefficients are known, the more accurate is the equation of state obtained from the virial expansion.[14, 16] Since the virial expansion converges slowly, commonly several types of approximants have been used in order to accelerate the convergence, and the constructed analytical EOSs give good results when compared with computer simulation data for the compressibility factor. The computer simulation data provided by Kolafa and Rottner[10] for reduced densities from 0.4 to 0.89 are currently considered to be the best available, and so are used as referents for the construction or testing of new expressions.[12-15]

Finally, it is important to bear in mind when proposing new EOSs or comparing EOSs is that the geometric properties of the hard-disk molecules demand that the analytical expressions should diverge at the closest packing fraction[18] $y_0 = \pi\sqrt{3}/6 = 0.90689....$. In addition, recently Santos and López de Haro have studied the idea that the radius of convergence of the virial series could be dictated by a branch-point singularity.[15]

Most of the expressions for the HD EOSs proposed in the literature were constructed with the approximation $y_0 = 1$, and thus cannot reproduce the singularity point.[13] Examples of these simple expressions (having less than six parameters) are the well-known Scaled Particle Theory[5] and Henderson[6,7] EOSs, with a proposal of Mulero *et al.*[14], denominated SMC2, being the most recent and accurate. Other simple expressions are the so-called SHY EOS[8] which gives the correct $y_0$ value, the Rusanov-4 EOS[9] with the pole being located at $y$ greater than 1, and two EOSs (KP1,

Eq. (12) in Ref. 12; KP2, Eq. (16) in Ref. 12) proposed by Khanpour and Parsafar[12] with poles at $y$ below 1. A recent comparison of the behaviour of these expressions when contrasted with computer simulation values for the compressibility factor[10] has been made by Mulero *et al.*[14]. With respect to the reproduction of virial coefficients, only KP1 and KP2 were constructed using recently calculated virial coefficient values, so they are the only simple expressions reproducing the fourth and fifth virial coefficients, respectively.

There are also some six-parameter expressions which can thus reproduce the values of the first seven virial coefficients. In particular, Santos and López de Haro[15] have recently shown that the rescaled-virial-expansion (RVE),[19] the [3,3] Padé approximant,[20] and their own branch-point-approximant (BPA) constructed using the values of the first seven virial coefficients can also reproduce approximately the values for the eighth to the tenth.

Of more complex expressions (more than 10 parameters involved), the most recent and successful proposals are those of Clisby and McCoy,[11] of Kolafa and Rottner,[10] and of Mulero *et al.*[14] In particular, Clisby and McCoy[11] proposed a [4, 5] and a [5, 4] Padé approximant (henceforth CM45 and CM54):

$$z_{CM45} = \frac{1 + 0.69939247(2y) - 0.33033017(2y)^2 + 0.11294457(2y)^3 - 0.012320562(2y)^4}{A_1}$$

(3)

with

$$A_1 = 1 - 0.30060752(2y) - 0.81172709(2y)^2 + 0.62751627(2y)^3 \\ - 0.17862580(2y)^4 + 0.021359218(2y)^5$$

$$Z_{CM54} = \frac{A_2}{1 - 1.0628945(2y) + 0.41940485(2y)^2 - 0.11367848(2y)^3 + 0.021846467(2y)^4}$$

(4)

with

$$A_2 = 1 - 0.062894522(2y) + 0.13851476(2y)^2 + 0.0067699403(2y)^3$$
$$+ 0.0039942056(2y)^4 + 0.00047760798(2y)^5$$

These EOSs contain 9 adjustable coefficients that were obtained by using the first ten virial coefficients. They each diverge at two real positive poles: 0.9722604325 and 1.059731986 for CM45, and 0.9477290073 and 1.140697980 for CM54.

Almost simultaneously, Kolafa and Rottner[10] proposed three analytical expressions (henceforth KR1, KR2, and KR3) in the form

$$Z_{KR} = \sum_i C_i \left(\frac{y}{1-y}\right)^i$$

(5)

which contain 12 or 13 $C_i$ parameters obtained from the first five virial coefficients and a fit to their own computer simulation results. They proposed these three different expressions in order to reproduce both the stable and metastable phases up to a maximum reduced density of 0.88 (KR1), 0.89 (KR2), and 0.9 (KR3). As mentioned in the introduction, these EOSs are the most accurate ones in reproducing the computer simulation data. This is because of the large number of parameters used, and because those same simulation data were used to construct the equations. Nevertheless they do not give the appropriate value for the closest-packing fraction, because they diverge at $y = 1$.

Finally, Mulero *et al.*[14] have proposed the SMC1 EOS:

$$z_{SMC1} = \frac{a_0^2 + \sum_{i=1}^{9} a_i y^i + a_{15} y^{15} + a_{20} y^{20} + a_{25} y^{25}}{(a_0 - y)^2} \quad (6)$$

which contains 13 parameters obtained using the exact (for $B_2$ to $B_4$) or numeric (for $B_5$ to $B_{10}$) values of the first ten virial coefficients from Ref. 11 and fitting to the computer simulation results of Ref. 10. The SMC1 EOS is more accurate than the CM ones in the reproduction of computer simulation values of $Z$ for high densities, although slightly less accurate than the KR ones. As can be seen, the pole of the SMC1 EOS is at $y = a_0 \approx 0.8391$, which is below the appropriate $y_0$ value.

The behaviour of these complex and accurate EOSs in prediction of virial coefficients higher than the tenth has yet to be considered. Table 1 presents the predicted values obtained for the eleventh to the eighteenth virial coefficients for the CM45, CM54, KR1, KR2, KR3, and SMC1 EOSs, together with the estimated values accepted by Clisby and McCoy,[11] where the uncertainty is in the last digit. The results, excluding the ones for SMC1, are also shown in Fig. 1, where we have included also predicted values from EOSs for $B_{19}$ and $B_{20}$. As can be seen, both CM45 and CM54 give accurate values for the $B_{11}$ to $B_{14}$ coefficients, but values outside the uncertainty for the higher ones with the exception of the very adequate value given by CM45 for $B_{18}$, as shown in Fig. 1. Figure 1 also shows the status of $B_{19}<B_{18}<B_{20}$ by CM45, which seems unreasonable. The KR EOSs give slightly worse values for all the coefficients, with the exception of the KR1 values for $B_{14}$ and $B_{15}$. For the higher coefficients, these EOSs give excessively low values when compared with those estimated by Clisby and McCoy.[11]

Finally in Table 1, one can see that the new SMC1 EOS, despite giving excellent values for the first ten virial coefficients, can only reproduce accurately the $B_{13}$ and $B_{14}$ coefficients, giving extremely low values for the higher coefficients and even negative values for $B_{17}$ and $B_{18}$.

In conclusion, none of the previous EOSs can reproduce the estimated values for $B_{11}$ to $B_{18}$ within the uncertainties of the estimated values of Clisby and McCoy.[11] In the case of the KR equations, this may be due to the fact that only five virial coefficients were used to construct the expression. In the case of the CM and SMC1 EOSs, most of their parameters were obtained from the first ten virial coefficients, but the use of approximate values for those parameters can lead to deviations in the prediction of higher virial coefficients. It would therefore be interesting to know if there is a method to construct new, but not more complex, EOSs giving those values.

In the three-dimensional case at least, the asymptotic expansion method proposed by Khanpour and Parsafar[12] has been shown[16] to be a very adequate method of constructing an accurate EOS starting only from the knowledge of the values of virial coefficients, without using any fitting procedure.

In the next section, we will explain the main bases of the method, and how we have extended and applied it in order to find new expressions for the HD EOSs.

## 3. New Asymptotic Expanded Method (AEM) EOSs

Recently, Khanpour and Parsafar[12] have proposed the asymptotic expansion method as a simple way to generate various EOSs in a unifying way in which the virial coefficients are used as reference. In particular, they developed several EOSs for the HD fluid by using the values of the first four and five virial coefficients. The proposed EOSs reproduce the computer simulation data for the compressibility factor with moderate accuracy at intermediate densities, but cannot be applied to the high-density or metastable ranges. Due to their simplicity, neither can the proposed EOSs reproduce higher virial coefficients.

Here we extend the AEM by using the first ten HD virial coefficients as reference and also by considering both positive and negative integer exponents in the expansion terms,[16] whereas only zero or negative values were considered by Khanpour and Parsafar[12]. Hence we assume that the HD EOS can be written as

$$Z_{AEM}(i,j) = \sum_{k=i}^{j} a_k (y - b_{ij})^{-k} \qquad j > i; i, j, k \in N \qquad (7)$$

where $a_k$ are coefficients to be determined, $b_{ij}$ is the radius of convergence of the virial expansion, and $k$ can take both positive and negative integer values.

The following step is to consider some constraints in order to select the appropriate number of coefficients and convergence radius. These constraints are: **(i)** consistency between the numerically calculated virial coefficients and the computer simulation data for the compressibility factor; **(ii)** accuracy is preferred to simplicity; **(iii)** the radius of convergence must be $b_{ij} > y_0$.

**(i) *Consistency.*** There are two ways to check the consistency of the numerically calculated virial coefficients with the computer simulation data for the compressibility factor. The first is that, if we include more accurate virial coefficients in the virial equations of state, they should approach the computer simulation data more accurately. The question then is what would be the appropriate order for the virial equation to reproduce the computer simulation data. To answer this question, we considered the virial EOS, Eq. (1), using the first ten virial coefficients given by Clisby and McCoy.[11] We thus generated ten virial EOSs, Eq. (1), with $i$ from 2 to 10. Then we compared the $Z$ values with the computer simulation data given by Erpenbeck and Luban[21] and those given by Kolafa and Rottner[10] for two different density ranges: the stable density range from 0.0385 to 0.70 (14 data points), and over the whole density range from 0.0385 to 0.89 (26 data points). Table 2 lists the absolute average deviations (AAD, %) for every virial EOS and density range. One observes in the table that the AAD obviously decreases with increasing virial order. When $i \geq 8$, the AAD in the region $\rho \in [0.0385, 0.70]$ approaches a value of less than 1%. Naturally, every equation yields a higher AAD value for the whole density region. Even with ten virial coefficients, the virial EOS cannot reproduce the whole density region data with an AAD below 2%.

The other way to check the consistency would be that the computer simulation data should yield correct virial coefficients. Because the published computer simulation data for a single method does not contain the sufficient detail, we could not do this check in the present work.

**(ii) *Accuracy is preferred to simplicity.*** In Eq. (7), the number of variables is *(j-i+1)* if *i>0*, and *(j-i+2)* if *i ≤ 0* (note that $j \geq 1$, see the bottom of point (*iii*) below). Expanding Eq. (7) and setting each virial coefficient equal to the numeric values, one can obtain the variables $a_k$ and $b_{ij}$. Because only the first ten virial coefficients have been exactly or numerically calculated, the number of variables ranges from 1 to 10. If we take it to be 10, then the first ten virial coefficients are obtained, but the resulting EOSs cannot be at all simple, and the high density region may not be described properly (as seen in Table 2 and Figure 2).

(iii) **The radius of convergence must be** $b_{ij} > y_0$. For hard-disk fluids, the closest packing fraction is expected to be in the equation of state, so that Eq. (7) must be used with $j \geq 1$.

In accordance with the above three constraints, we obtained 57 possible EOSs in the form of Eq. (7). For 56 of them, $j$ varies from 1 to 7 and $i$ may be one of several positive or negative integers. The last EOS is that obtained using $i = 0$ and $j = 8$.

From a careful check of these 57 EOSs, a significant number of options were rejected because complex $b_{ij}$ solutions appear, the seventeen and eighteenth virial coefficient are too large or too small compared with the values estimated by Clisby and McCoy [11], or $b_{ij} < y_0$. We finally considered only three well-behaved EOSs, as follows: $Z_{AEM}(i=-5, j=2)$, with $b_{-5,2} = 1.007201766$; $Z_{AEM}(i=-6, j=2)$, with $b_{-6,2}=$ 1.004590319; and $Z_{AEM}(i=-4, j=4)$, with $b_{-4,4}= 1.061330772$. The first one uses the exact or numeric values for the first nine virial coefficients, whereas the other two use the first ten. All of them have a $b_{ij}$ value slightly greater than 1. The $a_k$ values for these three EOSs, which we simply called *Z(-5,2), Z(-6,2)* and *Z(-4,4)*, are given in Table 3. These three equations are chosen for their ability to yield the exact, the numeric and the estimated virial coefficients and the appropriate computer simulation data for *Z*. Details are given in the following section.

## 4. Results

From the results in Table 2, one observes that: (*i*) some EOSs reproduce the first ten virial coefficients values, but cannot accurately describe the high density region; and (*ii*) for some others the case is the contrary, *i.e.*, they are constructed to accurately reproduce the compressibility factor values even at high densities, but give clearly wrong values for the higher-order virial coefficients.

The other aspect to be into account is the value of the packing fraction at which the EOS has a pole. We shall first consider these three aspects separately.

### 4.1 Accuracy in reproducing the compressibility factor

All the simple and complex HD EOSs mentioned in the Introduction can reproduce the *Z* computer simulation values at low densities. We therefore consider here only the results given by Kolafa and Rottner[10] in the density range from 0.4 to 0.89. In Table 4, the AAD between those values and the ones given by different simple and complex EOSs, ordered in terms of approximate analytical complexity, are given for three density ranges. Thus, AAD3 refers to the density range from 0.4 to 0.75 (8 data points), AAD4 to the highest densities from 0.8 to 0.89 (8 data points), and AAD5 to the whole density range from 0.4 to 0.89 (16 data points). In Figure 2, the data for the high density region are shown together with the predictions from different EOSs.

As mentioned above, the SMC2 expression of Mulero *et al.* [14] gives very accurate results over the whole density range, being the most accurate simple expression. The KPs, BPA, RVE, and Padé[3,3] expressions can be considered as of intermediate complexity. Of these four expressions, RVE gives the lowest AAD value for the lowest density range considered (AAD3), and KP2 the lowest for the high densities (AAD4). Naturally, the more complex SMC1 and KR expressions give excellent results (see Figure 2). Finally, the three AEM EOSs proposed here, which are slightly simpler than the CMs, the SMC1, and the KRs, give a similar accuracy to that of the CMs since they were both constructed using the values of the first ten virial coefficients.

We must stress that these three AEM EOSs were not chosen for giving the lowest AAD values when compared with computer simulation results for $Z$. Indeed, comparing the $Z(-5,2)$, $Z(-6,2)$, and $Z(-4,4)$ expressions with those of Khanpour and Parsafar [12], which are both equivalent to $Z_{AEM}(0,3)$ in Eq. (7) but use a value for $b$ which is fixed ($b=1.1$ in KP1) or calculated ($b=1.1363$ in KP2), one can see that the KP expressions give lower AAD values at high densities (AAD4), and therefore also over the whole density range (AAD5). This could be due to the high $b$ values they used. In fact, following the KP method, when the $Z_{AEM}(0,4)$ expression is considered (five virial coefficients known) the $b$ values obtained are lower than 0.77. With $Z_{AEM}(0,5)$ the AAD5 value for the whole density range increases to 1.46%, and with $Z_{AEM}(0,7)$ it decreases to 1.07% (similar to that obtained with KP1). For $Z_{AEM}(0,8)$ the $b$ value obtained is below 0.5, and thus clearly inadequate.

As a first conclusion, there seems to be that the behaviour of EOSs at high densities is more influenced by the *b* value than by the number of virial coefficients used to build them.

One can also try to construct new AEM EOSs in which the $a_k$ coefficients are obtained from the numeric virial coefficients, while the $b_{ij}$ value is obtained from a fit to the *Z* computer simulation data. In this case a $Z_{AEM}(i,j,b_{ij})$ EOS is constructed using one less virial coefficient than in the case of a $Z_{AEM}(i,j)$ EOS. Thus, 81 EOSs can be constructed, with $1 \leq j \leq 9$ (for *j=9*, there is only one choice *i=0)*. By focusing only on those new expressions giving an AAD5 value (for the whole density range) clearly below 1% (indeed the KP1 EOS is a $Z_{AEM}(0,3, 1.1)$ EOS, and it gives AAD5 = 1.1%), and including a $b_{ij}$ value between 0.9069 and 1.5, we found that the most accurate expressions are: (i) $Z_{AEM}(-2,5, 0.9073)$, which reproduces the exact or numeric values the first eight virial coefficients, has an excellent *b* value, and gives AAD5 = 0.61%; and (ii) $Z_{AEM}(0,8,1.0017)$, which reproduces the first nine virial coefficients, has a slightly poorer $b_{ij}$ value, but gives AAD5 = 0.54% and has only zero or positive values for *k* in Eq. (7), as do the KP expressions. The $a_k$ coefficients for these two expressions can be obtained easily, and are available upon request to the authors.

In sum, we have shown that the AEM expressions, as deduced only by knowing the first ten virial coefficients, cannot reproduce computer simulation data for *Z* more accurately than some other complex (KR1, for instance) or even less complex (SMC2, for instance) EOSs, as shown in Table 4 and Figure 2. Indeed, the AEM expressions give results that are very similar to those of the Clisby and McCoy Padé approximants, which were also constructed using the values of the first ten virial coefficients.

It has also been shown that, if one takes $b_{ij}$ in the AEM equation, Eq. (7), to be an adjustable parameter obtained from a fit to the computer simulation $Z$ data, the accuracy can be clearly improved, with a limiting value for the AAD over the whole density range of 0.54%, although still greater than that obtained with other expressions studied here (see Table 4).

The above conclusions are in agreement with those obtained recently by Mulero *et al.* [14], and also complete the results reported by Khanpour and Parsafar,[12] because they indicate the limiting behaviour of the AEM expressions when the only source used to construct them are the exact or numeric values for the virial coefficients. Of course, the method could give better results when more virial coefficients have been calculated. In any case, the aim of the present study was not to directly find EOSs giving a very low AAD, but rather ones that provide good predictions even for the estimated values of the first eighteen virial coefficients. This is the subject of the next subsection.

4.2 **Reproducing the eleventh to the eighteenth virial coefficients**

As is seen in Table 1, none of the EOSs that are most accurate in reproducing $Z$ values can reproduce the estimated values for $B_{11}$ to $B_{18}$ within the uncertainties of the estimated values of Clisby and McCoy,[11] despite reproducing adequately the values up to $B_{10}$.

In the preceding subsection, we considered five new equations as the best representatives of the AEM method. Of them, $Z_{AEM}(-2,5,b=0.9073)$ and $Z_{AEM}(0,8,b=1.0017)$ do not reproduce any virial coefficient higher than $B_8$ and $B_9$, respectively. The $B_{11}$ to $B_{18}$ values obtained from the other three ($Z(-5,2)$, $Z(-6,2)$, and $Z(-4,4)$) are given in Table 5. The $B_{11}$ to $B_{20}$ values from $Z(-4,4)$ are shown in Figure 1.

As can be seen, these three EOSs give adequate predictions for the $B_{11}$ to $B_{14}$ virial coefficients, with behaviour similar to that obtained with the CM45 and CM54 Padé approximants, and improving the results given by the SMC1 and KR expressions (see Table 1). For $B_{15}$ to $B_{18}$, the $Z(-6,2)$ EOS gives better results than $Z(-5,2)$, but neither is as accurate as $Z(-4,4)$. Indeed the $Z(-4,4)$ expression is the only published EOS giving accurate values for all the numerically calculated or estimated virial coefficients for hard disks.

A shortcoming of $Z(-4,4)$, however, is that its $b$ value is larger than the closest packing value 0.9069 for hard-disk fluids, and larger than that given by other EOSs. Nevertheless, we found that taking $b$ to be an adjustable parameter did not improve the accuracy in reproducing the virial coefficients. If in the future more accurate virials become numerically calculated, work on improving the present results would be interesting. In any case, it seems interesting to study the behaviour of the $b$ values that were obtained when the AEM procedure was used. This is done in the next subsection.

### 4.3 The distribution of the closest packing fractions

Figure 3 shows the real $b_{ij}$ values obtained for each of the 57 EOSs constructed using only values of the virial coefficients. Because the coefficients $a_k$ and pole $b_{ij}$ are obtained by solving equations with high integer-degree in $b_{ij}$, there may be several real solutions for the same equation of state.

In the case of $j = 1$, a decreasing function $b_{i1}(i)$ is obtained. In particular, only for $i = -6$ or $-7$ are values close to $y_0$ obtained. In particular, we found $b_{-7, 1} = 0.9076$, *i.e.*, slightly higher than $y_0$. This means that one cannot find a better $b_{i1}$ value by including more terms in this particular form of EOS, *i.e.*, using lower values of $i$, which needs the use of higher virial coefficients.

For $j = 7$, the tendency is not at all clear. But for $j=2$ to 6, one can see alternating tendencies as the value of $i$ changes. Moreover, some of the trends approach $y_0$ with decreasing values of $i$ ($i$ being more negative and thus increasing the number of virial coefficients used for the calculation).

In view of their capacity to reproduce the virial coefficients and the compressibility factor in the current context of the first ten virial coefficients, it would seem that the choice of $j = 2$ or 4 in the AEM equations might be the best way to construct new HD EOSs when further numeric values of higher virial coefficients than the tenth become calculated in the future.

## 5. Conclusions

In this paper, we have first shown that the most accurate currently available expressions for the hard-disk fluid cannot reproduce the estimated values of the eleventh to the eighteenth virial coefficients.

We then extended the asymptotic expansion method proposed by Khanpour and Parsafar to construct new HD EOSs by considering the currently available exact or numeric values for the first ten virial coefficients. This procedure led to 57 new expressions which were then checked for their reproduction of computer simulation values for the compressibility factor, of the higher virial coefficients, and of the closest packing limit of the packing factor, *i.e.*, the value of the convergence radius.

We found that only one of all the EOSs including those previously proposed and those newly proposed here can reproduce accurately the estimated values of the $B_{11}$ to $B_{18}$ coefficients. This was the expression denominated $Z(-4,4)$, which is Eq. (7) with $i = -4$ and $j = 4$, with $b_{-4,4} = 1.061330772$, and with the $a_k$ values given in Table 3. This EOS deviates from the computer simulation $Z$ values by an average value of 1.6% over the whole density range, which is an adequate value when compared with other expressions with a similar number of parameters and with expressions constructed using only virial coefficient values (hence, without using a fitting procedure). Although a shortcoming of $Z(-4,4)$ is that its $b$ value is larger than the closest packing value (0.9069), we showed that, when the $b$ value is taken to be an adjustable parameter in order to improve agreement with the $Z$ computer simulation results (reducing the deviation to only 0.54%), the accuracy in reproducing the virial coefficients is poorer.

We have also studied how the *b* value changes in the proposed 57 equations as the number of coefficients used increases. The results showed that the use of the particular values $j = 2$ or 4 in Eq. (7) could be the best way in order to construct new HD AEM EOSs when more accurate values of higher virial coefficients than the tenth become calculated in the future.

The present results constitute a complement to recently published studies, in particular to Refs. 1 and 12-14, and at the same time represent a departure point from which to study the performance of the AEM method in higher dimensions, *i.e.*, for hard-hypersphere fluids. Work in this direction is currently in progress.

## Acknowledgements

The National Natural Science Foundation of China under Grant No. 10804061, the Natural Science Foundation of Shandong Province under Grant No. Y2006A06, and the foundations of QFNU and DUT have supported this work (J.T. and Y.G.). A.M. thanks the Ministerio de Educación y Ciencia of Spain for support through Project FIS2006-02794 FEDER.

## Figure Captions

**Fig. 1** The plot of virial coefficients from the eleventh to the twentieth *vs.* the number

of the virial coefficients.

**Fig. 2** The plot of the compressibility factor $Z$ vs. the packing fraction $y$, for $y \geq 0.5$.

**Fig. 3** Real $b_{ij}$ values distributions for the 57 AEM EOSs constructed by using Eq. (7). The horizontal line represents the value $b_{ij} = y_0 = \pi\sqrt{3}/6$.

Figure 1

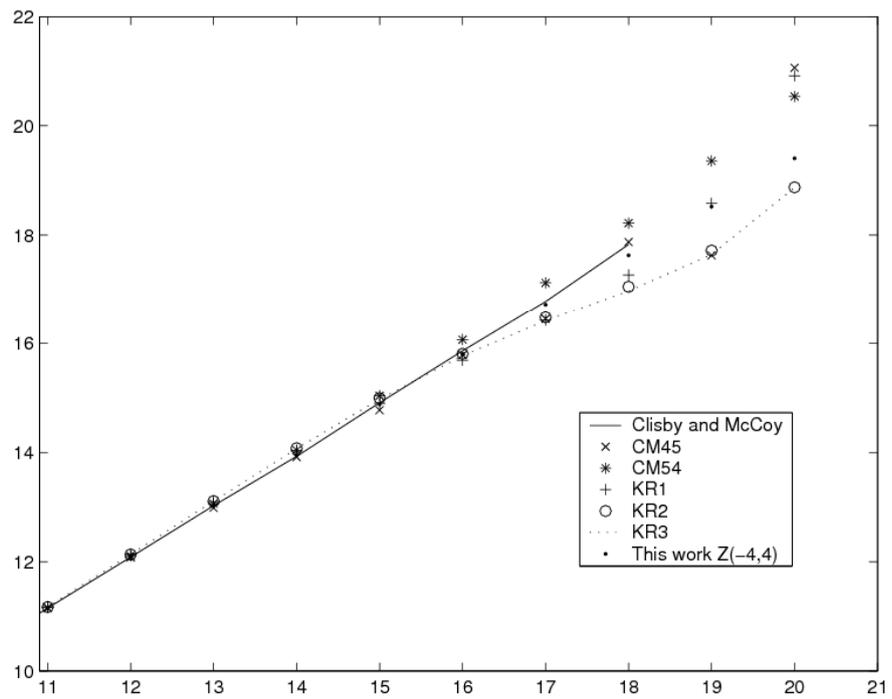

Figure 2

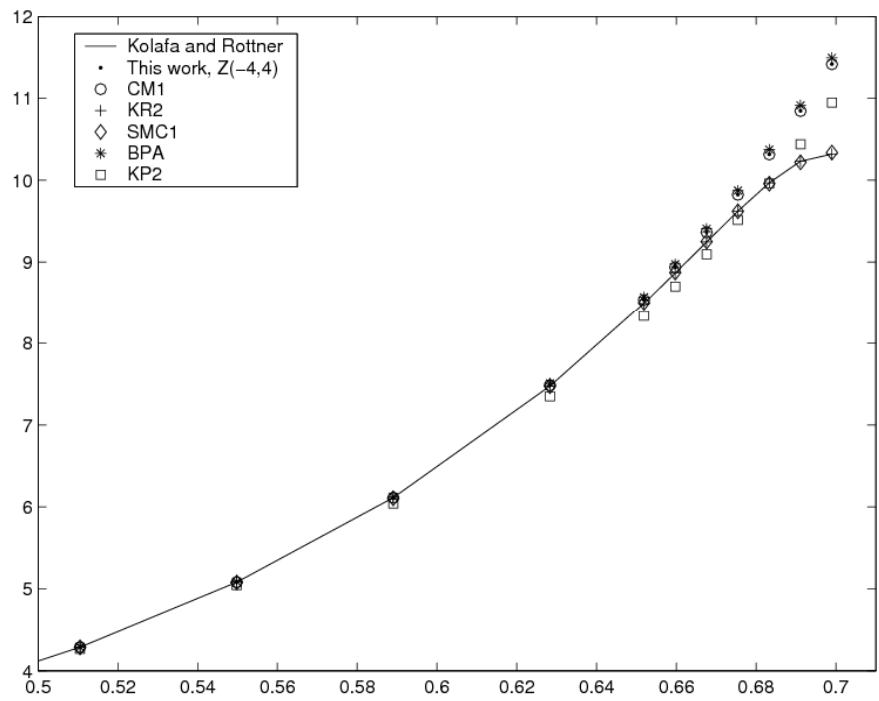

Figure 3

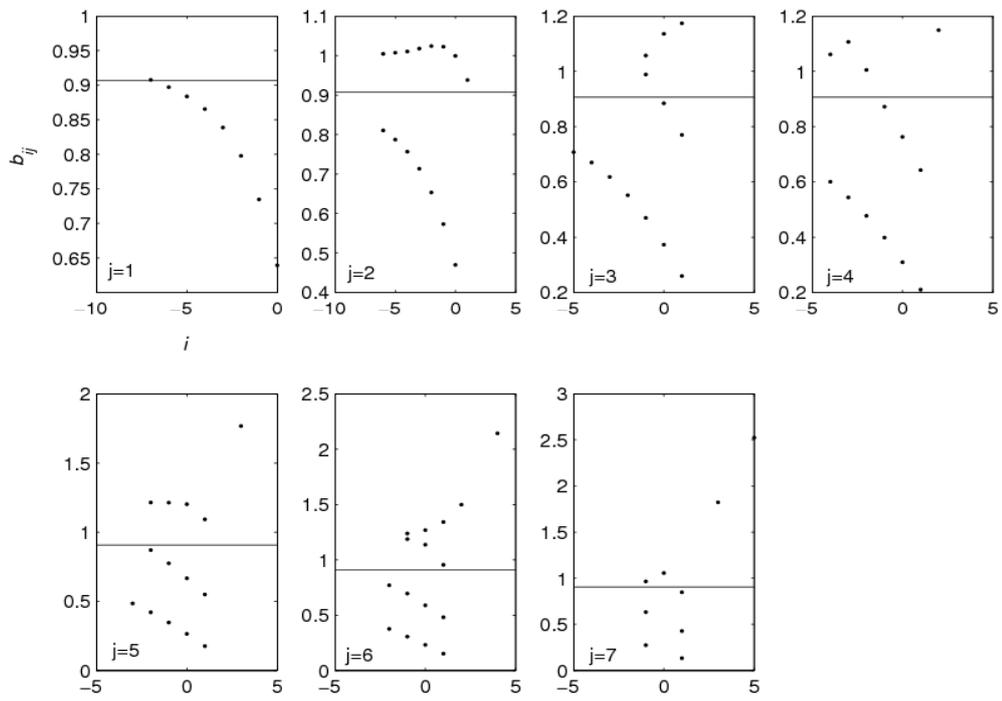

**Table 1.** Estimated values for eleventh to eighteenth virial coefficients for hard-disk fluids.

| Virial coefficients | Ref. 11 | CM54 | CM45 | KR1 | KR2 | KR3 | SMC1 |
|---|---|---|---|---|---|---|---|
| $B_{11}/B_2^{10}$ ($10^{-2}$) | 1.089 | 1.0896.. | 1.0886… | 1.0909… | 1.0909… | 1.0909… | 1.0919… |
| $B_{12}/B_2^{11}$ ($10^{-3}$) | 5.90 | 5.910… | 5.897… | 5.922… | 5.925… | 5.925… | 5.923… |
| $B_{13}/B_2^{12}$ ($10^{-3}$) | 3.18 | 3.189… | 3.169… | 3.195… | 3.200… | 3.200… | 3.187… |
| $B_{14}/B_2^{13}$ ($10^{-3}$) | 1.70 | 1.713… | 1.698… | 1.712… | 1.718… | 1.717… | 1.694… |
| $B_{15}/B_2^{14}$ ($10^{-4}$) | 9.10 | 9.178… | 9.017… | 9.093… | 9.152… | 9.140… | 8.872… |
| $B_{16}/B_2^{15}$ ($10^{-4}$) | 4.84 | 4.902… | 4.816… | 4.787… | 4.823… | 4.812… | 0.094… |
| $B_{17}/B_2^{16}$ ($10^{-4}$) | 2.56 | 2.612… | 2.509… | 2.506… | 2.514… | 2.504… | -3.037… |
| $B_{18}/B_2^{17}$ ($10^{-4}$) | 1.36 | 1.389… | 1.363… | 1.317… | 1.300… | 1.294… | -3.653… |

**Table 2** Absolute average percentage deviations (AAD, %) of virial equations with accurate virial coefficients compared with the computer simulation data [10, 21]. AAD1: data in the range from $\rho = 0.0385$ to $\rho = 0.70$, 14 data points; AAD2: range from $\rho = 0.0385$ to $\rho = 0.89$, 26 data points.

|  | $Z=1$ | $Z_2$ | $Z_3$ | $Z_4$ | $Z_5$ | $Z_6$ | $Z_7$ | $Z_8$ | $Z_9$ | $Z_{10}$ |
|---|---|---|---|---|---|---|---|---|---|---|
| **AAD1 (%)** | 52.74 | 30.02 | 16.85 | 9.30 | 5.09 | 2.78 | 1.52 | 0.83 | 0.45 | 0.25 |
| **AAD2 (%)** | 68.83 | 49.28 | 34.86 | 24.37 | 16.89 | 11.61 | 7.90 | 5.28 | 3.44 | 2.14 |

**Table 3** Coefficients for the proposed AEM EOSs, Eq. (7).

**Coefficients of *Z(-5,2)***

| k   | -5            | -4            | -3           | -2           |
|-----|---------------|---------------|--------------|--------------|
| $a_k$ | -0.007291184282 | -0.07394506681 | 0.3294073306 | -0.8294107225 |
| k   | -1            | 0             | 1            | 2            |
| $a_k$ | 1.229989976   | -0.9067074009 | -0.1700929215 | 1.007201766  |

**Coefficients of *Z(-6,2)***

| k   | -6             | -5             | -4            | -3            | -2           |
|-----|----------------|----------------|---------------|---------------|--------------|
| $a_k$ | -0.005263140845 | -0.05489758653 | -0.2652536306 | -0.7776081071 | -1.504014698 |
| k   | -1             | 0              | 1             | 2             |              |
| $a_k$ | -1.906843865   | -1.360599603   | -0.3675351823 | 1.037493293   |              |

**Coefficients of *Z(-4,4)***

| k   | -4             | -3             | -2            | -1             | 0            |
|-----|----------------|----------------|---------------|----------------|--------------|
| $a_k$ | -0.005546618059 | -0.04307289065 | -0.1230076654 | -0.08602246075 | 0.3728308751 |
| k   | 1              | 2              | 3             | 4              |              |
| $a_k$ | 0.8211504520   | 1.574755450    | 0.02177304185 | 0.03028727367  |              |

**Table 4** Absolute average deviations (AADs) of the compressibility factor from EOSs (ordered approximately in terms of mathematical complexity) when compared with the computer simulation results of Kolafa and Rottner[10] over different density ranges: AAD3 ($\rho$=0.4-0.75), AAD4 ($\rho$=0.8-0.89), AAD5($\rho$=0.4-0.89). The Rusanov equation is that for $n$=4 and $k$=0.876677 as given in Ref. 9. SH, RVE, and Padé[3/3] are equations (2), (5), and (6), respectively, of Ref. 15.

| EOS | Reference | *b* | AAD3 (%) | AAD4 (%) | AAD5 (%) |
|---|---|---|---|---|---|
| SHY | 8 | 0.9069 | 1.30 | 4.36 | 2.83 |
| SMC2 | 14 | 1 | 0.12 | 0.086 | 0.10 |
| Rusanov | 9 | 1.140670965 | 0.35 | 2.01 | 1.18 |
| KP1 | 12 | 1.10 | 0.060 | 2.17 | 1.11 |
| KP2 | 12 | 1.1363 | 0.37 | 2.05 | 1.21 |
| BPA | 15 | 1 | 0.049 | 3.63 | 1.84 |
| RVE | 15 | 1 | 0.024 | 3.38 | 1.70 |
| Padé [3/3] | 15 | 0.9208123809 1.319456660 2.254069305 | 0.033 | 3.74 | 1.89 |
| Z(-5,2) | This work | 1.007201766 | 3.81E-3 | 3.06 | 1.53 |
| Z(-4,4) | This work | 1.061330772 | 1.74E-3 | 3.13 | 1.57 |
| Z(-6,2) | This work | 1.004590319 | 1.61E-3 | 3.11 | 1.55 |
| CM1 | 11 | 0.9722604325 | 1.67E-3 | 3.11 | 1.56 |
| CM2 | 11 | 0.9477290073 | 2.85E-3 | 3.22 | 1.61 |
| SMC1 | 14 | 0.83908560190 | 3.92E-2 | 8.78 E-2 | 6.35 E-2 |
| KR1 | 10 | 1 | 9.50E-5 | 0.02 | 0.01 |
| KR2 | 10 | 1 | 1.31E-4 | 5.60 E-4 | 3.46 E-4 |
| KR3 | 10 | 1 | 2.02E-4 | 2.90 E-3 | 1.55 E-3 |

**Table 5** Estimated values of the eleventh to eighteenth virial coefficients for hard-disk fluids.

| Virial Coefficients | Ref. 11 | This work, Eq. (7) | | |
|---|---|---|---|---|
| | | $(i, j)= (-5,2)$ | $(-6,2)$ | $(-4,4)$ |
| | | $b = 1.007201766$ | $b = 1.004590319$ | $b = 1.061330772$ |
| $B_{11}/B_2^{10}$ ($10^{-2}$) | 1.089 | 1.0875… | 1.0890… | 1.0894… |
| $B_{12}/B_2^{11}$ ($10^{-3}$) | 5.90 | 5.883... | 5.898… | 5.904… |
| $B_{13}/B_2^{12}$ ($10^{-3}$) | 3.18 | 3.160… | 3.172… | 3.179… |
| $B_{14}/B_2^{13}$ ($10^{-3}$) | 1.70 | 1.688… | 1.697… | 1.703… |
| $B_{15}/B_2^{14}$ ($10^{-4}$) | 9.10 | 8.972… | 9.036… | 9.083… |
| $B_{16}/B_2^{15}$ ($10^{-4}$) | 4.84 | 4.748… | 4.790… | 4.823… |
| $B_{17}/B_2^{16}$ ($10^{-4}$) | 2.56 | 2.502… | 2.530… | 2.551… |
| $B_{18}/B_2^{17}$ ($10^{-4}$) | 1.36 | 1.314… | 1.331… | 1.344… |